\documentclass[runningheads]{llncs}
\usepackage[T1]{fontenc}
\usepackage{graphicx}
\usepackage{booktabs}
\usepackage{caption}
\usepackage{makecell}
\usepackage[misc]{ifsym}
\usepackage{bbding}
\usepackage[most]{tcolorbox}
\usepackage{amsmath}
\usepackage{multirow}
\usepackage[table,xcdraw]{xcolor}
\usepackage{diagbox}
\usepackage{subcaption}
\usepackage[most]{tcolorbox}
\usepackage{makecell}

\newcommand{\name}{LLM-SrcLog }

\begin{document}

\title{LLM-SrcLog: Towards Proactive and Unified Log Template Extraction via Large Language Models}


\author{Jiaqi Sun\inst{1\thanks{This work is done during his internship at Alibaba Group.}}\email{jotaro@sjtu.edu.cn} \and
Wei Li\inst{2}\email{jianhao@taobao.com} \and
Heng Zhang\inst{2}\email{zh383316@alibaba-inc.com} \and
Chutong Ding\inst{1}\email{dingct2004@tongji.edu.cn} \and
Shiyou Qian\inst{1,\thanks{Corresponding author.}}  \email{qshiyou@sjtu.edu.cn} \and
Jian Cao\inst{1}\email{cao-jian@sjtu.edu.cn} \and
Guangtao Xue\inst{1}\email{gt\_xue@sjtu.edu.cn}}

\authorrunning{J. Sun et al.}

\institute{Shanghai Jiao Tong University, School of Computer Science, 
800 Dongchuan RD, Shanghai 200240, China 
\and 
Alibaba Group,
969 West Wen Yi Road, Hangzhou 311121, China
}

\maketitle

\begin{abstract}
Log parsing transforms raw logs into structured templates containing constants and variables. It underpins anomaly detection, failure diagnosis, and other AIOps tasks. Current parsers are mostly reactive and log-centric. They only infer templates from logs, mostly overlooking the source code. This restricts their capacity to grasp dynamic log structures or adjust to evolving systems. Moreover, per-log LLM inference is too costly for practical deployment.
In this paper, we propose LLM-SrcLog, a proactive and unified framework for log template parsing. It extracts templates directly from source code prior to deployment and supplements them with data-driven parsing for logs without available code. LLM-SrcLog integrates a cross-function static code analyzer to reconstruct meaningful logging contexts, an LLM-based white-box template extractor with post-processing to distinguish constants from variables, and a black-box template extractor that incorporates data-driven clustering for remaining unmatched logs. Experiments on two public benchmarks (Hadoop and Zookeeper) and a large-scale industrial system (Sunfire-Compute) show that, compared to two LLM-based baselines, \name improves average F1-score by 2–17\% and 8–35\%. Meanwhile, its online parsing latency is comparable to data-driven methods and about 1,000$\times$ faster than per-log LLM parsing. LLM-SrcLog achieves a near-ideal balance between speed and accuracy. Finally, we further validate the effectiveness of LLM-SrcLog through practical case studies in a real-world production environment.

\end{abstract}

\section{Introduction}

Logs are a key telemetry channel for modern software systems, supporting critical operation tasks such as anomaly detection, failure diagnosis, and behavior auditing \cite{180271}. However, raw log messages are noisy and heterogeneous: they blend fixed text constants (e.g., error keywords or component names) with dynamic variables like IDs, parameters, and exception messages in diverse formats. Log parsing addresses this by converting raw logs into structured templates, which are then used to group events, build statistical models, and enable downstream analysis~\cite{logparser1,10025560}. The quality of these templates directly determines the effectiveness of system observability and AIOps pipelines \cite{10.1007/s10664-022-10214-6}.

Existing log parsing research largely follows three lines. Unsupervised clustering methods such as Drain~\cite{drain}, Spell~\cite{10.1145/1629575.1629587}, IPLoM~\cite{iplom}, and Logram~\cite{logram} discover templates by grouping structurally similar log messages, requiring no labels but relying on token-level heuristics. Machine Learning(ML)/Deep Learning(DL)-based methods like DeepLog~\cite{deeplog}, LogBERT~\cite{logbert}, and NuLog~\cite{nulog} encode logs as sequences and learn representations for anomaly detection or template inference. More recently, Large Language Model(LLM)-based approaches~\cite{loggpt,logparserllm,divlog,llmlog} exploit LLMs to generate templates directly from log texts, often using in-context learning or multi-round prompting.

Despite these advances, current methods share two major limitations. First, they are reactive and log-centric: templates are derived only after logs are produced, and most methods overlook the source code that actually defines logging behavior. This makes it hard to handle dynamically constructed messages, multi-function string composition, and exception-based logs, while also delaying the availability of high-quality templates. Second, even LLM-based parsers typically process isolated log messages as input. This leads to ambiguous separation between variables and constants, high inference cost, and limited generalization.

To overcome these limitations, we propose LLM-SrcLog, a proactive and code-oriented framework for log template parsing. Instead of waiting for logs to be generated at runtime, \name analyzes source code before deployment. It reconstructs the effective logging context via static analysis, and uses an LLM to extract semantically meaningful templates from this context. 
For logs with unavailable source code (e.g., third-party libraries or dynamically loaded components), \name incorporates an online unsupervised data-driven clustering method (Drain3~\cite{drain3}), thereby unifying white-box and black-box log parsing. A lightweight post-processing step further refines and validates the LLM outputs to ensure template consistency and usefulness. The code implementation is available at \textit{https://github.com/ottoSJTU/LLM-SrcLog}.

In summary, this paper makes the following contributions:  
(1) We propose LLM-SrcLog, a proactive log template extraction framework that combines static code analysis with LLMs to derive templates directly from source code rather than from log messages.  
(2) We conduct extensive experiments on two public Java benchmarks from LogPAI\cite{logparser1} (Hadoop\cite{hadoop} and Zookeeper\cite{zookeeper}) and a large-scale industrial monitoring application (Sunfire-Compute) from Alibaba Group, showing that compared to two LLM-based baselines (DivLog\cite{divlog} and LLMLog\cite{llmlog}), LLM-SrcLog improves average F1-score by 2–17\% and 8–35\% respectively, while achieving online parsing latency close to lightweight data-driven methods and roughly 1{,}000$\times$ faster than per-log LLM approaches.  
(3) We deploy LLM-SrcLog in real-world systems. It parses online logs and supports practical troubleshooting, further validating its effectiveness.

\section{Related Work}
Recent work has explored three main approaches: unsupervised clustering, traditional ML/DL models, and LLMs ~\cite{logparser1,logparser2}. Yet most of them operate reactively since they analyze logs after they are produced, without accessing the source code that generates them. This limits their ability to understand dynamic log structures or adapt to evolving systems.

\subsection{Unsupervised Clustering Methods}
Early rule-based methods like LogSig~\cite{logsig} and SLCT~\cite{slct} relied on manual definition of delimiters or templates, making them inflexible and expensive to maintain. Unsupervised clustering methods overcome this by automatically identifying templates from log streams using structural similarity, reducing the need for human intervention. Representative approaches include Drain~\cite{drain} which uses a tree to match token-aligned logs, Spell~\cite{spell} based on longest common subsequences, and IPLoM~\cite{iplom} which groups logs by field positions. These methods support scalable and deployable parsing in real-world systems.

However, these methods are primarily syntactic. They struggle to distinguish semantic constants (e.g., error codes) from variables (e.g., IDs), and cannot reconstruct logs generated dynamically from multiple function calls or exceptions. Moreover, they rely exclusively on runtime logs, disregarding the underlying source code that defines log origins. This limits their accuracy and adaptability in evolving software environments.

\subsection{Traditional ML/DL Methods}
To address the semantic blindness of syntactic clustering, natural language processing and DL models have been used to capture log semantics. These methods treat logs as token sequences and employ word embeddings, recurrent neural networks, or pre-trained language models to learn meaningful representations. Key examples are DeepLog~\cite{deeplog}, using LSTM \cite{lstm} to model normal log sequences for anomaly detection; LogBERT~\cite{logbert}, adapting BERT \cite{bert} for log encoding; NuLog~\cite{nulog}, framing parsing as masked language modeling to get embeddings for downstream tasks; and Logram~\cite{logram}, using n-gram dictionaries for efficient parsing. They shift focus from token matching to grasping log intent.

However, these models still miss contextual reasoning and cannot generalize to new logging styles without retraining or fine-tuning. Therefore, they struggle in cases needing deep program understanding, adaptive template generation, or cross-language log parsing—all key in modern, heterogeneous software systems.

\subsection{LLM-based Log Parsing}
LLMs are promising for log parsing due to strong program understanding and few-shot reasoning \cite{10.5555/3495724.3495883}. They interpret logging statements in rich code contexts, distinguish constants from variables across programming languages, and adjust to new log formats without retraining \cite{DBLP:journals/jnsm/KarlsenLZH24}. Recent work includes LogGPT~\cite{loggpt}, which uses GPT models \cite{openai2024gpt4technicalreport}  to analyse log semantics for anomaly detection. LogParser-LLM~\cite{logparserllm} leverages in-context learning for training-free parsing with adjustable granularity. LLMLog~\cite{llmlog} introduces a multi-round annotation framework with adaptive prompt selection to improve template accuracy. DivLog~\cite{divlog} selects diverse few-shot examples to guide LLM-based parsing through contextual reasoning. These studies demonstrate the potential of LLMs to surpass traditional data-driven methods in flexibility and interpretability.

While these approaches harness the generality of LLMs, they typically provide only isolated log messages or brief code snippets as input, which leads to high computational cost and slow processing speed. Also, Without inter-procedural context, LLMs cannot reliably reconstruct dynamically generated logs from helper functions or exception propagation. In short, these methods remain reactive and log-centric. They treat source code as an optional hint rather than the primary source of logging logic. Thus, they fail to support proactive template extraction from source code before deployment.

\section{Analysis and Observations}

\subsection{Proactive Template Discovery from Source Code}
Traditional log clustering methods, such as Drain~\cite{drain}, operate in a post-hoc manner that log messages must first be collected during system execution and then clustered.  This introduces inherent latency and often fails to capture anomalies that occur in the early stages of system operation, as their corresponding log templates may not yet have been observed.  In contrast, deriving log templates directly from source code is a proactive strategy \cite{10.5555/2387880.2387909,DBLP:conf/icic/SuLFZZ25}. It offers two major advantages: \textbf{timeliness} and \textbf{accuracy}.
\begin{itemize}
    \item \textbf{Timeliness} is achieved by analyzing code pre-deployment. Code-based extraction generates accurate templates from source logic, populating the template library before runtime. This enables immediate and accurate log matching from the first message.

    \item \textbf{Accuracy} is achieved by grounding templates in actual source code logic. This avoids the fragmentation common in data-driven methods, especially for complex logs involving conditionals, string concatenation, or embedded exceptions.

\end{itemize}



\subsection{Challenges Revealed by Real-World Log Observations}
While LLM-based log template extraction offers strong semantic comprehension via source code analysis, empirical observations from production logs in Alibaba Group’s internal monitoring system \textit{Sunfire} highlight practical limitations. In particular, two common types of logs remain inadequately handled by code-aware methods alone, even with sophisticated LLM reasoning.

First, numerous logs are dynamically constructed at runtime, particularly through exception embedding. For instance, log \texttt{ERROR DataController [...] - parse error:1d} is produced by template \texttt{logger.error(e.getMessage(), e)}, where message content originates from exception objects and lacks a literal representation within source code. Consequently, even ideal static analysis cannot reconstruct their templates.

Second, a significant portion of logs come from code-unavailable sources, such as third-party libraries, internal SDKs, or dynamically loaded modules. Their implementations are often absent from application’s source repository, like: 
\begin{itemize}
    \item \texttt{failed to update dom: <.*>} from \textit{vipserver.client.core.VIPClient} library
    \item \texttt{write exception <.*>} from \textit{metrics.reporter.bin.BinAppender} library
\end{itemize}
Since corresponding source code is unavailable, static analysis, whether augmented by LLMs or not, cannot generate templates for such messages, though they often signify critical system failures.

These findings highlight that although code-aware LLM parsing is powerful, it remains insufficient alone. In the following section, we tackle these challenges through specific framework designs.

\section{Framework}

\subsection{Design Principals}
Guided by the empirical findings above, we derive three core design principles for LLM-SrcLog:

\begin{enumerate}
\item \textbf{Context-aware semantic parsing for white-box logs}: Since dynamically constructed logs (e.g., exception messages) lack literal format strings in source code, template extraction must infer the intended logging behavior by analyzing cross-function invocations and expression semantics. 

\item \textbf{LLM-augmented variable
disambiguation}: Since log formats vary widely in syntax, natural language use, and placeholder styles, template extraction should generalize across styles without manual rules. We choose LLMs as the main extractor because they can robustly distinguish constants from variables across diverse and evolving logging templates.

\begin{figure}[tb]
\centering
\includegraphics[width=0.9\textwidth]{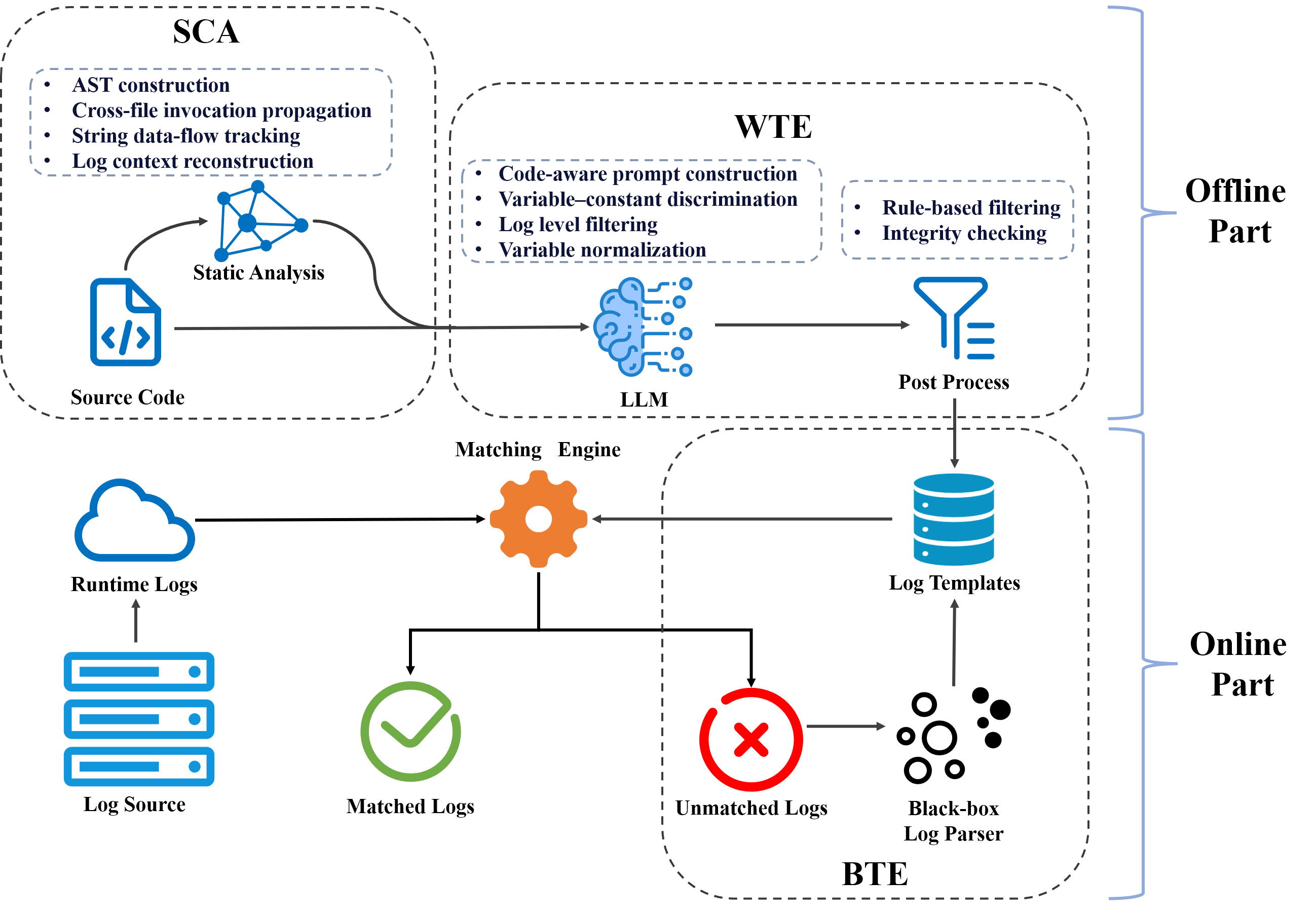}
\caption{System architecture of the \name framework. The pipeline integrates Static Code Analyzer (SCA), LLM-Driven White-box Template Extractor (WTE), and Black-box Template Extractor (BTE) to achieve comprehensive template coverage.}
\label{fig:framework}
\end{figure}

\item \textbf{Coverage for black-box logs}: For logs from components with unavailable code, neither static nor semantic analysis can recover true templates. Thus, the system must incorporate a fully autonomous, data-driven mechanism that operates without source knowledge, ensuring all logs are classified.
\end{enumerate}


\subsection{System Architecture}
The principles above lead to the following architecture, as illustrated in Figure~\ref{fig:framework}.

The first module, \textbf{Static Code Analyzer(SCA)}, constructs abstract syntax trees (ASTs) from source code across multiple programming languages and implements a cross-file context propagation mechanism. By tracing data flows across functions, it reconstructs the effective log content that appears at runtime. This context is then provided to the LLM, enabling it to generate meaningful templates for dynamically constructed logs, addressing the first design principle.

The second module, LLM-driven \textbf{White-box Template Extractor(WTE)} corresponds to the second design principle. It uses code context with log expressions as input and applies an LLM to detect constants and replace variables with standard placeholders. A post-processing step normalizes syntax, filters degenerate outputs, and deduplicates similar templates. 

The third module, \textbf{Black-box Template Extractor(BTE)}, embodies the third design principle. Logs that fail to match any template from the first two modules are routed to a data-driven parser (e.g., Drain3~\cite{drain3}), requiring no source code or human input. This module ensures both high fidelity for white-box logs and complete recall for black-box logs, fulfilling the dual objectives derived from real-world log observations.

\subsection{Static Code Analyzer (SCA)}

Traditional log template extraction methods assume that log messages are emitted via static format strings directly embedded in logging statements. However, modern software systems frequently construct log content dynamically through multiple function invocations, conditional string concatenation, or exception message propagation. 

To bridge this semantic gap, our static code analyzer module is designed not merely to locate logging calls, but to reconstruct effective logging contexts, which is the set of all code paths that could contribute to the final log messages. This reconstruction is achieved through an inter-procedural, path-sensitive analysis that traces data flows from the logging invocation back through the call graph to all possible return expressions in upstream helper functions.

Consider an illustrative example in the code boxes below. In \texttt{Foo.java}, the logging statements invoke static methods in \texttt{Bar.java} that conditionally construct message strings based on input parameters. A naive parser would see only method calls like \texttt{Bar.getUserName("0")} and fail to extract any meaningful template, as no format string is present at the caller side.

\begin{lstlisting}[caption={Foo.java},label=lst:foo,language=Java,backgroundcolor=\color{blue!5!white},frame=single,framesep=10pt,rulecolor=\color{blue!10!white},basicstyle=\ttfamily\small,keywordstyle=\color{blue}\bfseries,stringstyle=\color{red},commentstyle=\color{gray}\itshape,captionpos=b]
package com.example;
import com.example.Bar;

public class Foo {
    public void logSomething(String type) {
        if ("ERROR".equals(type)) {
            log.error(Bar.getUserName("0"));
        } else {
            log.fatal(Bar.getDisplayName("1"));
        }}}
\end{lstlisting}

\begin{lstlisting}[caption={Bar.java},label=lst:bar,language=Java,backgroundcolor=\color{blue!5!white},frame=single,framesep=10pt,rulecolor=\color{blue!10!white},basicstyle=\ttfamily\small,keywordstyle=\color{blue}\bfseries,stringstyle=\color{red},commentstyle=\color{gray}\itshape,captionpos=b]
package com.example;

public class Bar {
    public static String getUserName(String uid) {
        if (uid.startsWith("user")) {
            return "User_" + uid.toUpperCase() + "NotFound";
        } else {
            return "Invalid_User_ID" + uid;
        }}
    public static String getDisplayName(String uid) {
        if (uid.startsWith("guest")) {
            return "Guest_" + uid.toUpperCase();
        } else {
            return "Unknown_" + uid;
        }}}
\end{lstlisting}

The analyzer recursively resolves all function invocations and enumerates all feasible execution paths. For \texttt{Bar.getUserName("0")}, it identifies two branches: \texttt{uid.startsWith("user")} is true yielding \texttt{"User\_"} + \texttt{uid.toUpperCase()} + \texttt{"\_NotFound"} and as it is false yielding \texttt{"Invalid\_User\_ID\_"} + \texttt{uid}. Built-in methods like \texttt{toUpperCase()} are recognized and terminated with a marker, while the string literals and variable placeholders are preserved. This yields feasible templates  \texttt{"User\_<.*>\_NotFound"}, \texttt{"Invalid\_User\_ID<.*>"}, \texttt{"Guest\_<.*>"}, and \texttt{"Unknown\_<.*>"}. 

This path enumeration strategy explicitly addresses the challenge of dynamic string composition: a single logging statement may produce multiple distinct message templates depending on runtime conditions. Rather than forcing a single over-generalized template, our approach preserves this diversity by generating a set of candidate templates, each grounded in a specific control-flow path.

The final static analysis report of the example project is shown in Appendix ~\ref{lst:example-report}, which includes:
\begin{itemize}
    \item All feasible message templates derived from static paths;
    \item The functions source code leading to each template;
    \item Metadata indicating whether the message involves conditional logic or external method calls.
\end{itemize}


\subsection{LLM-driven White-box Template Extractor (WTE)}


The LLM-driven White-box Template Extractor extracts log templates available from source code, and addresses a fundamental limitation of traditional log parsing: the inability to interpret logs whose form diverges significantly from their source-level representation due to dynamic construction. Our solution bridges this gap by framing template extraction as a contextual reasoning task, where  LLM acts as an interpreter that synthesizes static program structure with observed runtime behavior.

Central to this design is the construction of a rich prompt that encodes both syntactic and semantic cues. Rather than feeding raw logs to the model in isolation, we condition the LLM on the structured static analysis report that captures cross-file call chains and data flows. By presenting this context alongside the original source code, we enable the LLM to distinguish between deterministic string literals and dynamic expressions, even when the latter originate from code not directly visible at the logging site.

The prompting strategy enforces three critical inductive biases to ensure robustness. First, it mandates strict placeholder normalization: all variables, function calls, and SLF4J-style \cite{slf4j2024} placeholders are  replaced with a uniformed representation. Second, it preserves original concatenation semantics, thereby maintaining fidelity to the actual log generation logic. Third, it restricts output to a machine-readable JSON schema that includes method identity, log level, and template string, facilitating downstream automation and validation. The full prompt specification, including input format, task instructions, and output schema, is provided in Appendix ~\ref{prompt:1}.

Despite the LLM’s contextual awareness, there are still two challenges: (1) degenerate outputs where templates collapse into all-wildcard templates, and (2) over-generalization in natural-language-style logs where fixed error phrases are misclassified as variables. To mitigate these, we introduce a lightweight post-processing module that combines rule-based filtering (e.g., discarding templates with insufficient static content) with semantic validation via a secondary LLM invocation. This verifier assesses whether a template meaningfully discriminates failure modes, ensuring that only semantically coherent templates enter the template repository.

This architecture transforms the LLM from a black-box generator into a context-aware interpreter grounded in program semantics. By integrating static analysis as prompt context rather than a standalone extractor, the system achieves zero-shot adaptability across logging styles while maintaining high precision without reliance on handcrafted rules or format-specific heuristics.

\subsection{Black-box Template Extractor (BTE)}
A critical challenge in large-scale log parsing arises from the inherent incompleteness of any single-source template discovery approach. As demonstrated in our empirical study, logs emitted by third-party libraries, runtime frameworks, or dynamically loaded components collectively referred to as \textit{black-box logs}, lacking corresponding source code in the application repository.

To ensure universal coverage without sacrificing fidelity, our framework introduces a hybrid template fusion mechanism that synergistically combines two orthogonal sources of template knowledge: (1) semantically precise templates derived from code-aware analysis and LLM interpretation, and (2) empirically discovered templates via online log parser of raw log streams. This duality directly addresses the dichotomy observed in production systems: white-box logs benefit from high-precision, context-grounded template extraction, while black-box logs require a fallback that operates purely on data regularities.

The data-driven component leverages a streaming-compatible, unsupervised black-box log parser (Drain3 \cite{drain3} in our implementation) that incrementally clusters unmatched logs based on token-level structural similarity. This module operates in a fully autonomous manner: it requires no access to source code, no human-defined rules, and no prior knowledge of logging conventions. It excels at capturing recurring templates in exception messages, library-generated diagnostics, and ad-hoc debug statements—precisely the categories that dominate the “unmatched” tail in real-world deployments.
By design, this module embraces a precision-recall trade-off: semantic templates maximize precision for while-box templates, while clustered templates maximize recall for black-box templates. 


\section{Experimental Evaluation}
This section presents experimental results on two public benchmarks (Hadoop \cite{hadoop} and Zookeeper \cite{zookeeper}) and an industrial production dataset (Sunfire-Compute). Across all backbone models (Qwen3-8B, Qwen3-32B, and Qwen3-Max) \cite{yang2025qwen3technicalreport}, LLM-SrcLog achieves better template extraction quality than current LLM-based and data-driven methods. It also maintains time cost similar to traditional data-driven approaches and is much more efficient than existing LLM-based parsers. We further deployed LLM-SrcLog in a real-world production environment and validated its effectiveness through practical case studies.

\subsection{Experimental Setup}

\subsubsection{Datasets}
Due to mature static analysis tools and common benchmark datasets being available for Java, we apply LLM-SrcLog on two Java-based systems from LogPAI benchmark~\cite{logparser1} (Hadoop and Zookeeper), along with logs from Alibaba’s internal monitoring tool \textit{Sunfire-Compute}. The static-analysis part is language-agnostic and can support other languages with added front ends. The statistics of these datasets are summarized in Table~\ref{tab:datasets}.

\begin{table}[tbp]
\centering
\renewcommand{\arraystretch}{1.0}
\caption{Statistics of evaluation datasets.\label{tab:datasets}}
\begin{tabular}{lccc}
\toprule
\textbf{Statistic } & \textbf{Hadoop} & \textbf{Zookeeper} & \textbf{Sunfire-Compute} \\
\midrule
\# Log files     & 1003 & 393 & 983 \\
\# Log messages  & 2000 & 2000 & 6433 \\
\# Log templates\footnotemark[1] & 99  & 47   & 9 \\
\# Source code templates\footnotemark[2] & 156  &  222  &  220 \\
\bottomrule
\end{tabular}
\vspace{-3mm}
\end{table}

\footnotetext[1]{The distinct templates matched by the sampled log messages in each dataset.}
\footnotetext[2]{The total number of templates in the entire codebase.}

\subsubsection{Evaluation Metrics}

We evaluate template quality using precision, recall, and F1-score at template level. Following prior work~\cite{logram,10.1145/1629575.1629587}, we adopt a strict correctness criterion, as proper separation of static text and dynamic variables is crucial for downstream log analysis (e.g., anomaly detection based on variable variations). Similar to Dai et al.~\cite{logram}, we inspect parsing results and regard a parsed template correct if and only if there is a ground-truth (GT) template with matching constant segments and variable placeholders. 
Any mistake in static text or variable bounds counts as an error.

To assess efficiency, we measure wall-clock time for online log parsing. Baseline methods perform streaming matching over incoming logs and generate templates on-the-fly. Our approach splits the process into two steps: (1) offline template extraction from source code, and (2) online log parsing via regular-expression-based streaming matching against the template set. For all methods, the online parsing time is averaged over 10 runs.

\subsubsection{Baselines}

We compare our LLM-SrcLog with five representative log parsers:

\begin{itemize}
    \item \textbf{Drain}~\cite{drain}: a widely used streaming log parser based on a fixed-depth parse tree and word-level similarity.
    \item \textbf{Spell}~\cite{10.1145/1629575.1629587}: an online parser that leverages longest common subsequences to extract templates.
    \item \textbf{Logram}~\cite{logram}: an n-gram-based log parsing method that builds dictionaries to efficiently generate templates.
    \item \textbf{DivLog}~\cite{divlog}: an LLM-based log parser that selects diverse few-shot examples to guide template generation.
    \item \textbf{LLMLog}~\cite{llmlog}: an LLM-driven multi-round annotation framework with adaptive prompt selection for log template extraction.
\end{itemize}

We conduct all experiments on a MacBook Pro equipped with an Apple M2 chip featuring 8 CPU cores and 16 GB unified memory, running macOS Monterey 12.5. For all baseline methods, we use the default hyperparameters provided from their official implementations. In LLM-SrcLog, we use Drain3 \cite{drain3} as the black-box parser and do static code analysis based on the javalang \cite{javalang} library.   
For LLM-based methods (DivLog, LLMLog, and LLM-SrcLog), we set temperature to 0.1 for table output ~\cite{SAC-KG}. To test the robustness of LLM-SrcLog with different backbone models, we use three backbones: \textbf{Qwen3-8B}, \textbf{Qwen3-32B}, and \textbf{Qwen3-Max}(larger than 1T) \cite{yang2025qwen3technicalreport,qwen3max}, and present results per setting.

\subsection{Template Quality Evaluation}

Table~\ref{tab:quality_metrics} shows template-level precision, recall, and F1 scores. Compared with DivLog, LLM-SrcLog improves F1 by around 2\%, 1\%, and 17\% on Hadoop, Zookeeper, and Sunfire-Compute, respectively (averaged across three backbone models). Compared with LLMLog, the average F1 gains are about 8\%, 11\%, and 35\%. Several observations can be drawn from Table~\ref{tab:quality_metrics}.

\begin{table*}[tbp]
\centering
\renewcommand{\arraystretch}{1.1}
\setlength{\tabcolsep}{5pt}
\caption{Template-level recall(R), precision(P), and F1 of tested log parsers on three datasets, grouped by backbone models.}

\label{tab:quality_metrics}
\scalebox{0.87}{
\begin{tabular}{l|ccc|ccc|ccc}
\toprule
\multirow{2}{*}{Method} &
\multicolumn{3}{c|}{\textbf{Hadoop}} &
\multicolumn{3}{c|}{\textbf{Zookeeper}} &
\multicolumn{3}{c}{\textbf{Sunfire-Compute}} \\
\cline{2-10}
& R & P & F1 & R & P & F1 & R & P & F1 \\
\midrule
Drain                & 0.263 & 0.252 & 0.257 & 0.533 & 0.522 & 0.527 & 0.333 & 0.231 & 0.273 \\
Logram               & 0.061 & 0.058 & 0.059 & 0.378 & 0.362 & 0.370 & 0.111 & 0.050 & 0.069 \\
Spell                & 0.141 & 0.084 & 0.105 & 0.422 & 0.333 & 0.373 & 0.003 & 0.111 & 0.006 \\
\midrule
DivLog (8B)    & 0.684 & 0.619 & 0.650 & 0.920 & 0.820 & 0.868 & 0.889 & 0.667 & 0.762 \\
LLMLog (8B)    & \textbf{0.842} & 0.479 & 0.611 & 0.820 & 0.719 & 0.766 & \textbf{1.000} & 0.224 & 0.366 \\
\name(8B)      & 0.717 & \textbf{0.696} & \textbf{0.707} & \textbf{0.932} & \textbf{0.821} & \textbf{0.872} & \textbf{1.000} & \textbf{1.000} & \textbf{1.000} \\
\midrule
DivLog (32B)   & 0.667 & 0.559 & 0.608 & 0.900 & \textbf{0.875} & \textbf{0.887} & \textbf{1.000} & 0.700 & 0.750 \\
LLMLog (32B)   & \textbf{0.826} & 0.537 & 0.651 & 0.840 & 0.794 & 0.816 & \textbf{1.000} & 0.667 & 0.800 \\
\name(32B)     & 0.667 & \textbf{0.647} & \textbf{0.657} & \textbf{0.955} & 0.824 & 0.884 & 0.889 & \textbf{0.889} & \textbf{0.889} \\
\midrule
DivLog (Max)   & 0.746 & \textbf{0.759} & \textbf{0.752} & 0.920 & \textbf{0.893} & 0.906 & \textbf{1.000} & 0.750 & 0.857 \\
LLMLog (Max)   & \textbf{0.842} & 0.441 & 0.579 & 0.840 & 0.724 & 0.778 & \textbf{1.000} & 0.500 & 0.667 \\
\name(Max)     & 0.717 & 0.703 & 0.710 & \textbf{0.977} & 0.878 & \textbf{0.925} & \textbf{1.000} & \textbf{1.000} & \textbf{1.000} \\
\bottomrule
\end{tabular}
}
\vspace{-4mm}
\end{table*}

1) LLM-based methods consistently outperform traditional parsers. Across all three datasets, Drain, Logram, and Spell achieve significantly lower precision and recall. 
This consistent gap highlights a fundamental capability difference: traditional parsers rely on word-level structure and simple heuristics, whereas LLMs understand logging semantics, distinguish constants from variables, and handle varied formats (natural language, JSON, concatenated strings, exception messages). Thus, LLM-based methods generate more accurate and unified templates.

2) LLM-SrcLog generally achieves the best template quality. 
The key reason is that LLM-SrcLog does not treat logs as isolated text. It uses source code and static analysis to construct candidate templates and call-chain context before invoking the LLM. Remaining black-box logs are further routed to Drain3 for processing, ensuring coverage. This code-based approach reduces both over-generalized and fragmented templates, yielding better precision and recall than log-focused LLM parsers.

3) Hadoop poses more challenging, especially for LLM-SrcLog.  
All methods achieve lower F1 on Hadoop than on Zookeeper and Sunfire-Compute. This aligns with our dataset analysis: Hadoop has more logs from code-unavailable parts or runtime content not recoverable from source. For these, LLM-SrcLog uses Drain3, which limits the quality of the template. Still, LLM-SrcLog outperforms traditional baselines, proving that code-aware semantic extraction brings significant gains even in tough cases.


\begin{table}[tbp]
\centering
\renewcommand{\arraystretch}{1.05}
\setlength{\tabcolsep}{6pt}
\caption{Offline template extraction time (in minutes) of LLM-SrcLog.}
\label{tab:offline_time}
\begin{tabular}{l|ccc}
\toprule
Base model      & Hadoop & Zookeeper & Sunfire-Compute \\
\midrule
Qwen3-8B   & 35.7 & 25.1 & 18.5 \\
Qwen3-32B  & 40.6 & 30.4 & 21.3 \\
Qwen3-Max  & 41.9 & 27.7 & 24.8 \\
\bottomrule
\end{tabular}
\vspace{-4mm}
\end{table}

\subsection{Efficiency Evaluation}

We next evaluate the time cost of LLM-SrcLog in both the offline source-code-based template extraction phase and the online log parsing phase.

\paragraph{Offline code-based template extraction.}
Table~\ref{tab:offline_time} shows the time LLM-SrcLog takes to extract templates from source code across datasets using different backbone models.
Results indicate that for Java systems with around 1{,}000 source files, LLM-SrcLog finishes template extraction in 20–40 minutes. This offline step is done before deployment, so it does not affect online log parsing latency. Once the template library is built, logs can be parsed by regular-expression matching without LLM calls.

\begin{table*}[ tbp]
\centering
\renewcommand{\arraystretch}{1.1}
\setlength{\tabcolsep}{5pt}
\caption{Online parsing time (seconds) for tested parsers on three datasets.}
\label{tab:online_time}
\begin{tabular}{l|ccc}
\toprule
Method &
\makecell{Hadoop\\(2000 logs)} &
\makecell{Zookeeper\\(2000 logs)} &
\makecell{Sunfire-Compute\\(6433 logs)} \\
\midrule
Drain              & 0.117   & 0.094   & 0.791   \\
Logram             & 0.305   & 0.054   & 8.831   \\
Spell              & 0.163   & 0.110   & 2.381   \\
\midrule
DivLog (Qwen3-8B)  & 1377.8  & 1254.8  & 8844.9  \\
DivLog (Qwen3-32B) & 2172.6  & 1717.0  & 14588.3 \\
DivLog (Qwen3-Max) & 3597.6  & 2529.7  & 9138.6  \\
\midrule
LLMLog (Qwen3-8B)  & 5009.4  & 1035.9  & 13949.6 \\
LLMLog (Qwen3-32B) & 1373.8  & 518.6   & 19672.2 \\
LLMLog (Qwen3-Max) & 1863.9  & 862.3   & 7595.5  \\
\midrule
\name(Qwen3-8B)    & 2.157   & 0.743   & 0.643   \\
\name (Qwen3-32B)   & 2.147   & 0.766   & 0.644   \\
\name(Qwen3-Max)   & 2.120   & 0.753   & 0.620   \\
\bottomrule
\end{tabular}
\vspace{-4mm}
\end{table*}

\paragraph{Online parsing efficiency.}
Table~\ref{tab:online_time} presents the time in seconds (s)  each method takes to parse the evaluation logs. The results highlight three key points.

1) Direct LLM-based parsing is too slow, making it impractical in production environments.  
DivLog and LLMLog use an LLM for each small log batch. Parsing 2{,}000 logs takes around 10\textsuperscript{3}s. 
Yet, real-world systems produce vast log amounts. Alibaba Cloud DBaaS, for instance, can generate nearly 50 million logs (about 100~GB) per second during working hours~\cite{logstore}. Scaling per-log LLM inference would lead to days or even weeks of delay to process one second of logs. 

2) LLM-SrcLog achieves near data-driven efficiency with significant better quality.  
Its online parsing time is comparable with Drain, Logram, and Spell, often only seconds for thousands of logs.
This speed of LLM-SrcLog comes from decoupling heavy semantic reasoning: the LLM works offline to produce a high-quality template library, while online parsing uses simple regular expression matching against this library, scaling slowly with template count.
Notably, LLM-SrcLog remains robust with base model scaling at runtime.  

LLM-SrcLog combines the strengths of LLM-based semantic understanding with traditional stream parsers. It achieves better precision and recall by proactively extracting templates from source code, while keeping online costs low via efficient template matching instead of log parsing with LLM inference.

\subsection{Ablation Study}

To quantify the impact of each component, we conduct an ablation study on all three datasets by disabling one module at a time, using Qwen3-Max as the backbone model. We choose Qwen3-Max because LLM-SrcLog exhibits stable performance across different backbone models. Table~\ref{tab:ablation} shows template-level recall(R), precision(P), F1, and the number of extracted templates(N) for each variant.
Overall, removing any module typically reduces template quality. However, the effects vary by module, showing their unique functions.

First, static code analysis is key for rich logging context. Without it, the LLM only sees local log statements and a weaker call-chain view. Thus, the model outputs more incomplete or too-general templates.  Generally, precision drops more than recall, indicating that without context, the LLM misclassifies constants as variables and produces less discriminative templates.

Second, the post-processing module serves dual roles: rescuing useful templates and filtering invalid ones. Without it, we observe a substantial recall drop on Hadoop (by 20.2\%) and a moderate F1 decrease on Zookeeper (by 8.8\%). This happens because malformed or low-quality templates from the LLM are not detected and retried. Also, invalid or meaningless templates are not removed, reducing precision. Thus, post-processing acts as a quality control that both recovers missing templates and blocks flawed ones.

Third, the black-box template extractor(BTE) module presents a trade-off. Without it, precision and F1 may rise. This occurs because unmatched logs are directed to BTE (Drain3~\cite{drain3}), which has lower template accuracy. Adding templates from Drain3 boosts coverage but brings more errors, reducing overall precision. This aligns with the template quality evaluation results on Hadoop, where many code-unavailable logs force greater reliance on Drain3. As shown in Table~\ref{tab:ablation}, removing the BTE module reduces the number of extracted templates on Hadoop from 101 to 72, which is the largest drop across all datasets, confirming that a substantial portion of Hadoop logs originate from black-box components that only the BTE module can capture. Nevertheless, this module is essential: omitting it drops all unmatched logs, causing systematic recall loss on black-box logs and impairing observability. We favor full coverage and tolerate a slight precision drop to ensure each log matches a template.

\begin{table*}[tbp]
\centering
\renewcommand{\arraystretch}{1.1}
\setlength{\tabcolsep}{5pt}
\caption{Ablation study of LLM-SrcLog on three datasets. Each variant removes one module from the full system. Best values per column are highlighted in bold. }
\label{tab:ablation}
\scalebox{0.76}{
\begin{tabular}{l|cccc|cccc|cccc}
\toprule
\multirow{2}{*}{Variant} &
\multicolumn{4}{c|}{Hadoop} &
\multicolumn{4}{c|}{Zookeeper} &
\multicolumn{4}{c}{Sunfire-Compute} \\
\cline{2-13}
& R & P & F1 & N & R & P & F1 & N & R & P & F1 & N \\
\midrule
w/o SCA                  & 0.677 & 0.670 & 0.673 & 100 & 0.932 & 0.788 & 0.854 & 52  & 0.888 & 0.800 & 0.842 & 10 \\
w/o post process         & 0.515 & 0.809 & 0.630 & 63  & 0.932 & 0.759 & 0.837 & 54  & 0.889 & 0.889 & 0.889 & 9  \\
w/o BTE                  & 0.636 & \textbf{0.875} & \textbf{0.737} & 72  & 0.932 & 0.872 & 0.901 & 47  & 0.889 & \textbf{1.000} & 0.941 & 8  \\
w/ all modules           & \textbf{0.717} & 0.703 & 0.710 & 101 & \textbf{0.977} & \textbf{0.878} & \textbf{0.925} & 49  & \textbf{1.000} & \textbf{1.000} & \textbf{1.000} & 9  \\
\bottomrule
\end{tabular}
}
\vspace{-4mm}
\end{table*}

\subsection{Deployment in Real Production Environments}

To validate the effectiveness of \name in real-world settings, we deployed it in two production applications within Alibaba Group and enabled real-time parsing for ERROR-level logs. The monitoring window spanned 24 hours from November 30, 2025 to December 1, 2025. Due to confidentiality policies, we anonymize the applications as \textit{App1} and \textit{App2}.

App1 is responsible for scheduling, executing, and managing the full lifecycle of metadata-related background tasks in the monitoring platform. App2 serves as a unified service interface layer, providing metadata management, monitoring data access, and system integration capabilities for both internal and external systems. Table~\ref{tab:prod_apps} summarizes the key characteristics of the two applications and their corresponding log parsing statistics.

We observed that App1 produces far more error-level logs than App2. A closer inspection of the parsed templates reveals that a single template
\texttt{\textbar TAIR\_ERROR\textbar <.*>\textbar} 
accounts for most of the errors in App1. On average, this template matches 15{,}056 error logs per minute, which corresponds to 99.3\% of all parsed error logs in App1 during the monitoring period. TAIR is a distributed fast-access memory (MDB) and persistent (LDB) storage service~\cite{tair}.
Guided by this dominant template, we first examined TAIR read/write hit ratios, which appeared normal. We then hypothesized that network issues might be responsible for the TAIR access failures. Network telemetry confirmed this hypothesis. The average network retransmission rate of App1 during the observation window was 35.8\%, whereas that of App2 was only 14.0\%. The elevated retransmission rate in App1 explains the frequent TAIR access errors.

This case study demonstrates that \name supports downstream root cause analysis by providing high-quality templates for online logs in production environments, and the pre-extracted template library enables low-latency matching.
The temporal relationships between error logs and network retransmissions in both applications are further illustrated in Appendix~\ref{fig:app_logs_network}.

\begin{table}[tbp]
\centering
\renewcommand{\arraystretch}{1.0}
\setlength{\tabcolsep}{6pt}
\caption{Statistics of the two production applications and online parsing results over 24 hours. \textit{QPS} stands for \textit{Queries Per Second}.}
\label{tab:prod_apps}
\begin{tabular}{lcccc}
\toprule
\textbf{Application} & \textbf{Machines} & \textbf{QPS} & \textbf{Templates} & \textbf{Avg. logs per minute} \\
\midrule
App1 & 40 & 1,590 & 16 & 15,159 \\
App2 & 91 & 7,608 & 3 & 344 \\
\bottomrule
\end{tabular}
\vspace{-7mm}
\end{table}

\section{Conclusion and Future Work}

We propose LLM-SrcLog, a proactive log template parser incorporating static code analysis, LLM-based template extraction, and data-driven template extraction. By extracting templates from source code before deployment and enhancing them via data-driven clustering for black-box logs, \name achieves higher precision and recall than traditional parsers and log-focused LLM-based methods, while keeping online parsing cost near lightweight baselines. Future work will extend static analysis to more programming languages and optimize the online matching strategy by partitioning the global template set into application-specific subsets derived from the codebase, thereby further reducing matching overhead while preserving accuracy.

\section{Acknowledgment}
This work was supported by Alibaba Group through Alibaba Innovative Research Program.

\bibliographystyle{splncs04}
\bibliography{ref}

\appendix
\section{Appendix}
\subsection{Detailed Prompt Templates}
\begin{tcolorbox}[
  title=\small{Prompt template for LLM-Based Log template Extraction},
  colback=blue!5!white,
  colframe=blue!10!white,
  coltitle=black,
  fonttitle=\bfseries,
  breakable=true,
  label = prompt:1,
]
You are an expert Java log template extractor. Please extract all log templates by given the source code and static analysis report.

\textbf{Input format:}
\begin{itemize}
    \item Java source code
    \item Static analysis report containing:
    \begin{itemize}
        \item Location, method name, and initial template for each log call
        \item Call paths (cross-method/cross-file), showing for each level:
        \begin{itemize}
            \item \texttt{Class}: fully qualified class name (e.g., \texttt{com.example.A})
            \item \texttt{Call code}: method invocation statement
            \item \texttt{Called function info}: The source code of the called function)
        \end{itemize}
    \end{itemize}
\end{itemize}

\textbf{Your task:}
\begin{itemize}
    \item For each log call and each of its paths, construct the final log template by concatenating string literals extracted from return statements across the call chain.
    \item Replace the following with \texttt{<.*>}:
    \begin{itemize}
        \item Unknown functions,
        \item Built-in methods,
        \item Variable names,
        \item Any non-literal string components,
        \item All \texttt{\{\}} placeholders in the original log statement (SLF4J style), regardless of their runtime value.
    \end{itemize}
    \item Preserve all deterministic string constants exactly as they appear.
    \item Output must be a JSON array. Each element has the format: \{"method": class\_path.method\_name, "template": constructed\_template, "level": log\_level\}
\end{itemize}

Now process the following input:
\begin{itemize}
    \item     java\_code: \{java\_code\}
    \item     static\_analysis\_report:
    \{static\_analysis\_report\}
\end{itemize}
\end{tcolorbox}

\subsection{Static Analysis Report of the Example Project}
\begin{lstlisting}[caption={Static analysis report for Foo.java \& Bar.java},label=lst:example-report,language=Java,backgroundcolor=\color{blue!5!white},frame=single,framesep=10pt,rulecolor=\color{blue!10!white},basicstyle=\ttfamily\small,keywordstyle=\color{blue}\bfseries,stringstyle=\color{red},commentstyle=\color{gray}\itshape,captionpos=b]
Extracted 2 log calls

=== Analysis of log call 1 ===
Location: line 8, method: error
Template: undefined template

=== Call path analysis results (2 paths in total) ===

--- Path 1 ---
  1. Class: com.example.Foo
     Call code: log.error(Bar.getUserName("0"))
     Callee information: log method invocation
  2. Class: com.example.Bar
     Call code: Bar.getUserName("0")
     Callee information:
       public static String getUserName(String uid) {
         if (uid.startsWith("user")) {
           return "User_" + uid.toUpperCase() + "_NotFound";
         } else {
           return "Invalid_User_ID_" + uid;
         }
       }
  3. Class: java.lang.String
     Call code: uid.toUpperCase()
     Callee information: built-in method

--- Path 2 ---
  1. Class: com.example.Foo
     Call code: log.error(Bar.getUserName("0"))
     Callee information: log method invocation
  2. Class: com.example.Bar
     Call code: Bar.getUserName("0")
     Callee information:
       public static String getUserName(String uid) {
         if (uid.startsWith("user")) {
           return "User_" + uid.toUpperCase() + "_NotFound";
         } else {
           return "Invalid_User_ID_" + uid;
         }
       }

=== Analysis of log call 2 ===
Location: line 10, method: fatal
Template: undefined template

=== Call path analysis results (2 paths in total) ===

--- Path 1 ---
  1. Class: com.example.Foo
     Call code: log.fatal(Bar.getDisplayName("1"))
     Callee information: log method invocation
  2. Class: com.example.Bar
     Call code: Bar.getDisplayName("1")
     Callee information:
       public static String getDisplayName(String uid) {
         if (uid.startsWith("guest")) {
           return "Guest_" + uid.toUpperCase();
         } else {
           return "Unknown_" + uid;
         }
       }
  3. Class: java.lang.String
     Call code: uid.toUpperCase()
     Callee information: built-in method

--- Path 2 ---
  1. Class: com.example.Foo
     Call code: log.fatal(Bar.getDisplayName("1"))
     Callee information: log method invocation
  2. Class: com.example.Bar
     Call code: Bar.getDisplayName("1")
     Callee information:
       public static String getDisplayName(String uid) {
         if (uid.startsWith("guest")) {
           return "Guest_" + uid.toUpperCase();
         } else {
           return "Unknown_" + uid;
         }
       }

A total of 2 log calls, with 4 complete paths found.
\end{lstlisting}

\subsection{Online Deployment Results: Error Logs and Network Retransmissions}
\label{fig:app_logs_network}

\begin{figure}[tbp]
\centering
\begin{subfigure}[b]{\textwidth}
\centering
\includegraphics[width=\textwidth]{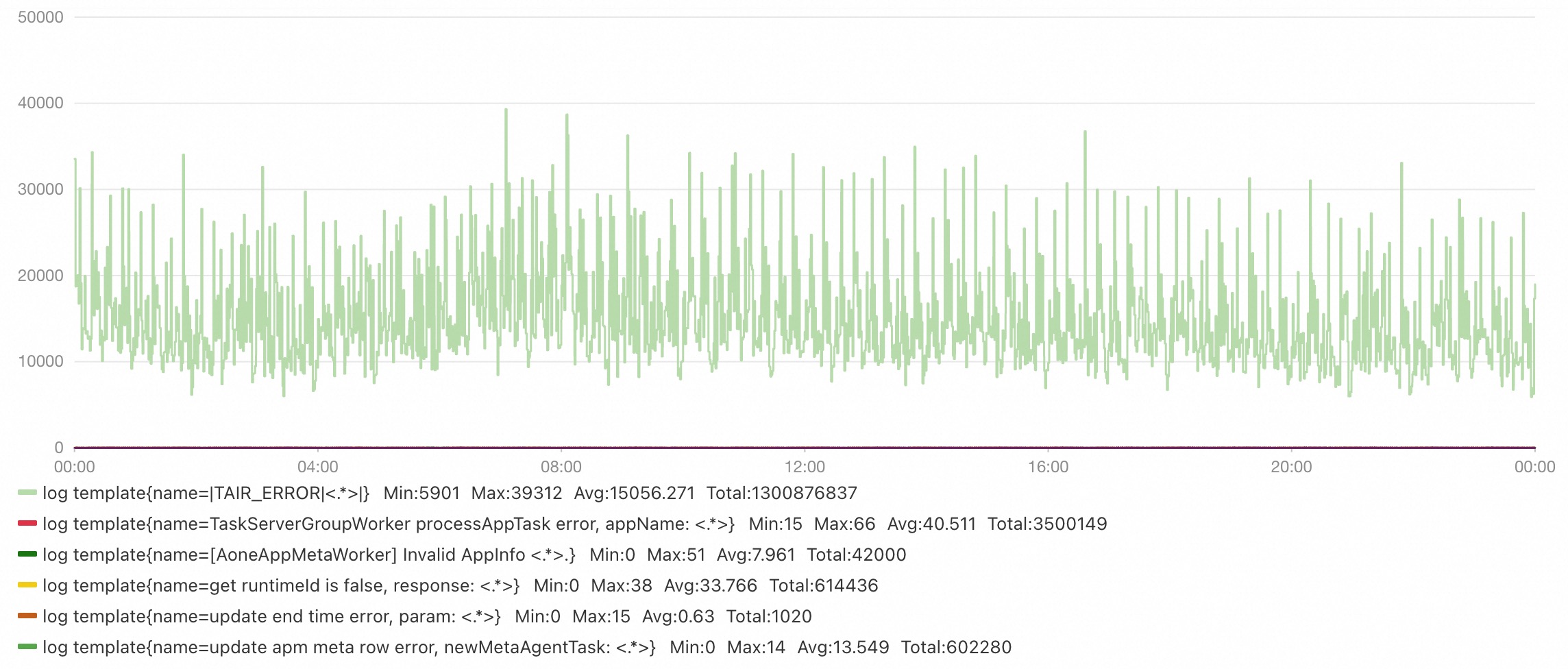}
\caption{App1 error logs over time.}
\label{fig:app1_logs}
\end{subfigure}

\vspace{2mm}

\begin{subfigure}[b]{\textwidth}
\centering
\includegraphics[width=\textwidth]{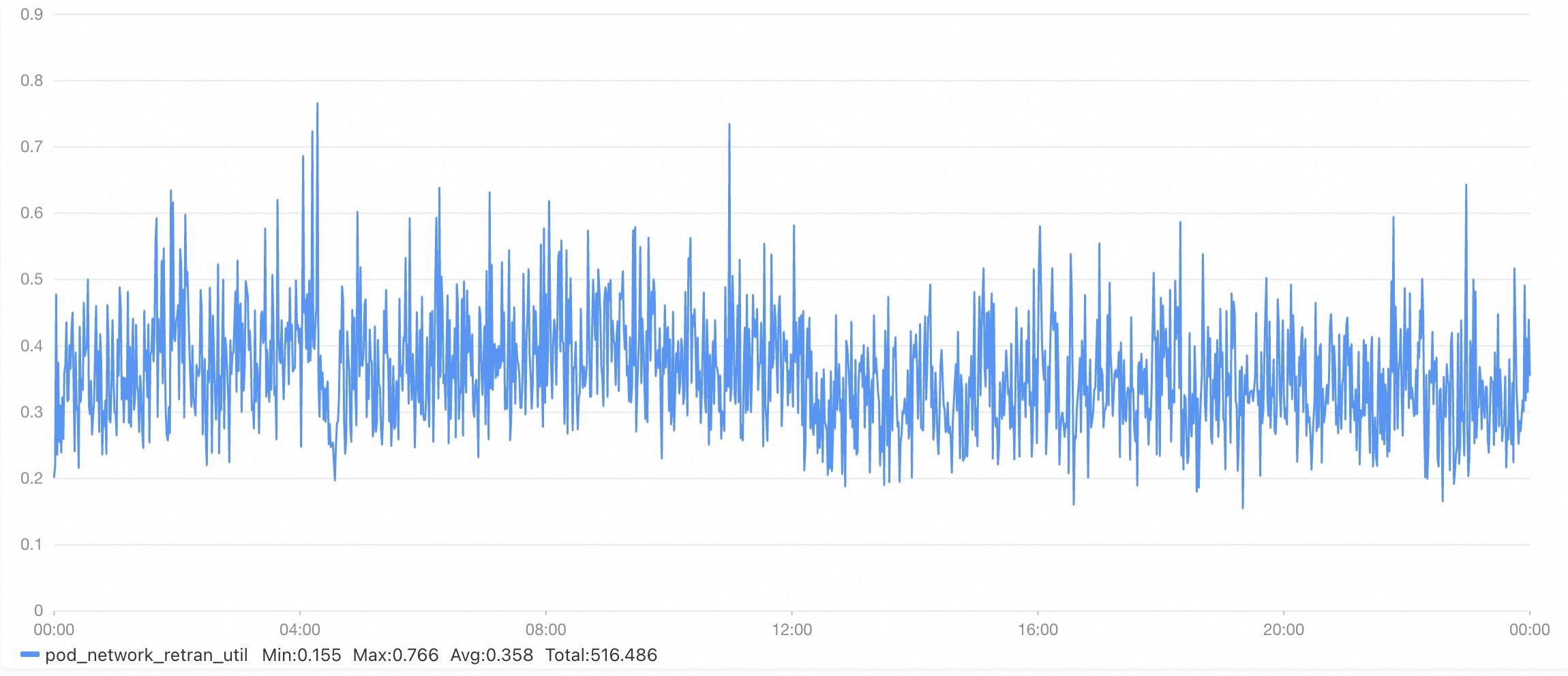}
\caption{App1 network retransmission rate over time.}
\label{fig:app1_network}
\end{subfigure}
\caption{Temporal trends of error logs and network retransmissions for App1 over a 24-hour period. In both subfigures, the horizontal axis denotes time within the observation window. In Subfigure~\subref{fig:app1_logs}, the vertical axis denotes the average number of error-level log messages per minute that are successfully matched to templates. In Subfigure~\subref{fig:app1_network}, the vertical axis denotes the average network retransmission rate per minute.}
\label{fig:app1_logs_network}
\vspace{-4mm}
\end{figure}

\begin{figure}[tbp]
\centering
\begin{subfigure}[b]{\textwidth}
\centering
\includegraphics[width=\textwidth]{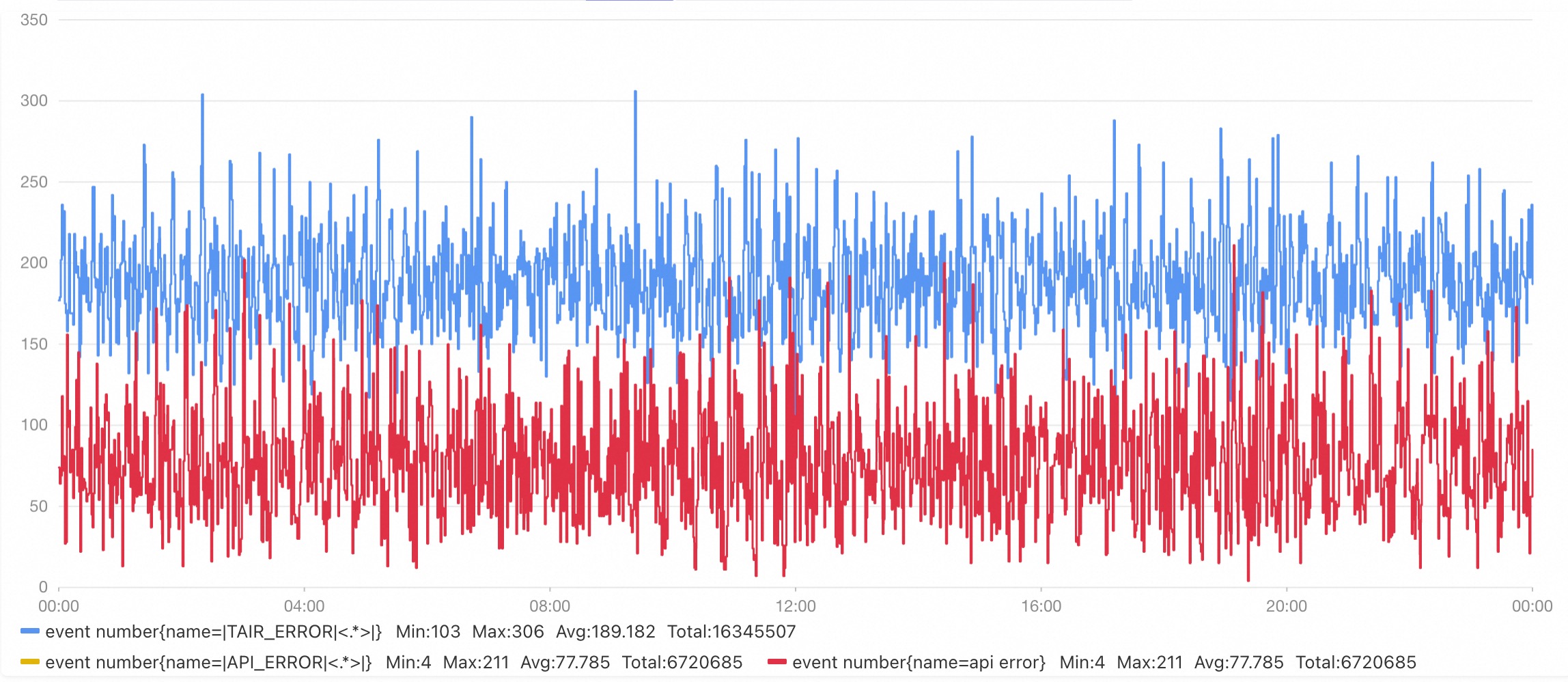}
\caption{App2 error logs over time.}
\label{fig:app2_logs}
\end{subfigure}

\vspace{2mm}

\begin{subfigure}[b]{\textwidth}
\centering
\includegraphics[width=\textwidth]{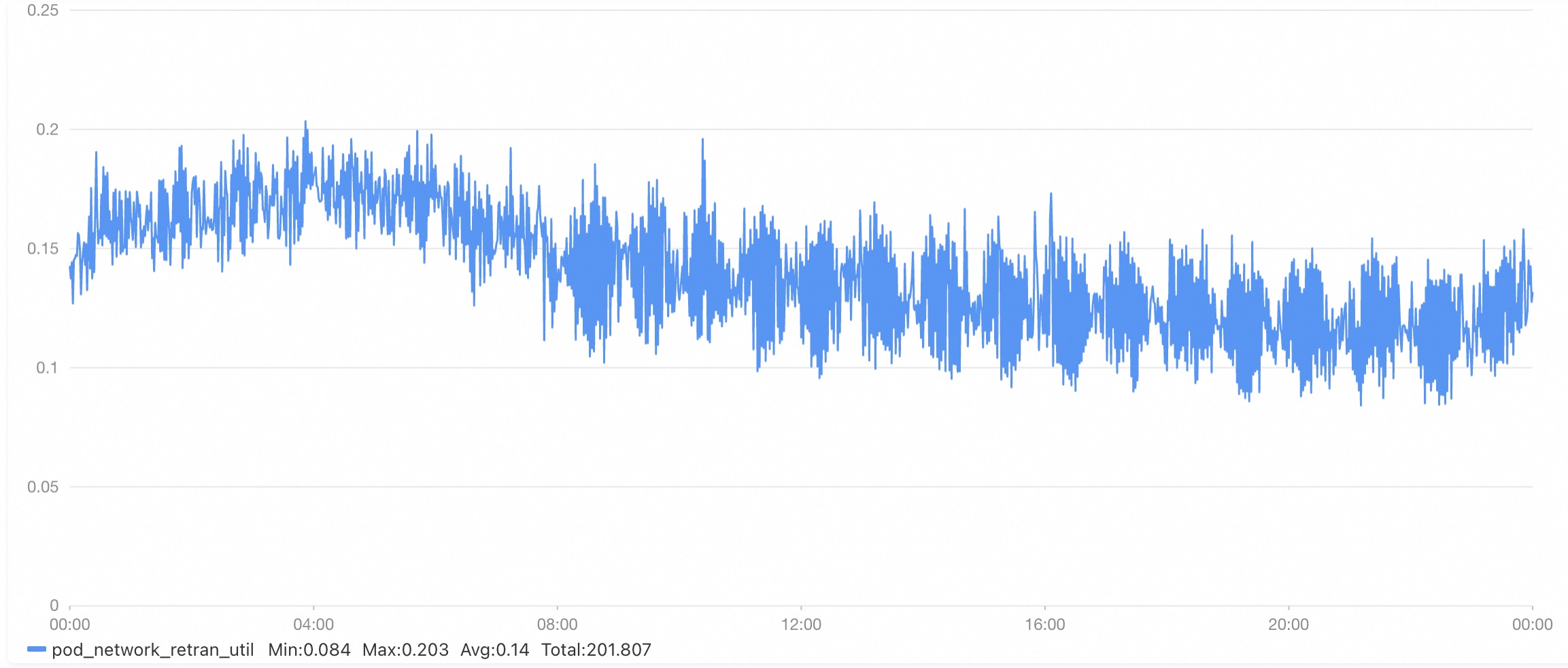}
\caption{App2 network retransmission rate over time.}
\label{fig:app2_network}
\end{subfigure}
\caption{Temporal trends of error logs and network retransmissions for App2 over a 24-hour period. In both subfigures, the horizontal axis denotes time within the observation window. In Subfigure~\subref{fig:app2_logs}, the vertical axis denotes the average number of error-level log messages per minute matched to templates. In Subfigure~\subref{fig:app2_network}, the vertical axis denotes the average network retransmission rate per minute.}
\label{fig:app2_logs_network}
\vspace{-4mm}
\end{figure}

\end{document}